# Crystalline ZnMgSe:Cr$^{2+}$: a new material for active elements of tunable IR-lasers

Zagoruiko Yu.A., Kovalenko N.O., Fedorenko O.A., Khristyan V.A.

*Institute for Single Crystals STC "Institute for Single Crystals" NAS of Ukraine*
*60, Lenin Ave., Kharkov, Ukraine*
*e-mail: zagoruiko@isc.kharkov.ua*

The obtaining of a new class of laser crystals to be used as active elements of tunable lasers working in medium IR-region (2…5 µm) is a topical problem for modern materials science. Such compounds are produced by doping of A$^{II}$B$^{VI}$-group single crystals and their solid solutions with Cr$^{2+}$ or Fe$^{2+}$ ions [1,2]. As shows comparative analysis of the optical and electrophysical characteristics of these materials and of the generation characteristics of the active elements made on their base, ZnSe:Cr$^{2+}$ crystals seem to be most promising from the viewpoint of their practical application. It should be noted that for the first time comprehensive investigations of the structural, mechanical, optical and electrophysical properties of Zn$_{1-x}$Mg$_x$Se single crystals (0.03≤x≤0.55) were performed in [3-5]. On the basis of the mentioned studies, it was proposed to use hexagonal crystals of Zn$_{1-x}$Mg$_x$Se (0.12≤x≤0.50) substitution solid solution as a thermostable material for polarization and electrooptical elements meant for near and medium IR regions [6].

Reported in the present work, which is a development of the investigations [3-5], are the optical characteristics of Zn$_{1-x}$Mg$_x$Se:Cr$^{2+}$ single crystals (x≈0.21). This material was obtained by diffusion doping of 10x10x1 mm$^3$ Zn$_{1-x}$Mg$_x$Se samples with chromium. Cr diffusion proceeded from the vapor phase at 1250 K during 120 hours in hermetically sealed quartz ampoules from which air was preliminarily evacuated.

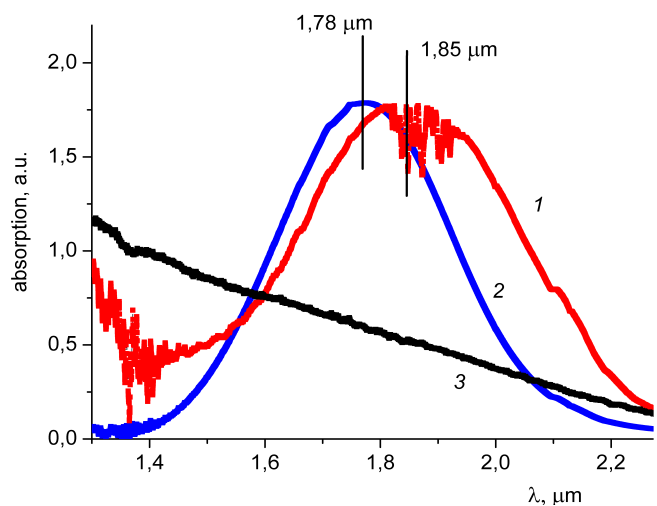

Fig. Absorption spectra of single crystal samples:
1- Zn$_{1-x}$Mg$_x$Se:Cr$^{2+}$ (x≈0.21), 2-ZnSe:Cr$^{2+}$,
3- Zn$_{1-x}$Mg$_x$Se (x≈0.21).

The optical transmission spectra of ZnSe:$Cr^{2+}$, $Zn_{1-x}Mg_xSe$ (x≈0.21) and $Zn_{1-x}Mg_xSe:Cr^{2+}$ (x≈0.21) samples were measured using the spectrometer Perkin Elmer Spectrum One FT-IR. Presented in the figure are the optical absorption spectra of these samples calculated from their optical transmission spectra taking into account the corresponding refraction coefficients. As is seen, the optical transmission spectra for ZnSe:$Cr^{2+}$ and $Zn_{1-x}Mg_xSe:Cr^{2+}$ (x≈0.21) samples have strong absorption bands with maxima at 1.780 μm and 1.855 μm, respectively. The presence of such absorption bands testifies to the fact that $Zn_{1-x}Mg_xSe:Cr^{2+}$ single crystals may be used as a new thermostable material for active elements of tunable lasers working in medium IR region, with a generation band shifted to longer wavelengths with respect to that of ZnSe:$Cr^{2+}$.

References

1. H. Jelinkova, P. Koranda, M.E. Doroshenko, T.T. Basiev, J. Sulc, M. Nemec, P. Cerny, V.K. Komar, and M.B. Kosmyna. $Cr^{2+}$:ZnSe laser pumped by 1.66μm or 1.97 μm radiation. Laser Phys. Lett. 1–7 (2006) / **DOI** 10.1002/lapl.200610065
2. Akimov V.A., Voronov A.A., Kozlovskiy V.I., Korostelin Yu.V., Landman A.I., Podmar'kov Yu.P., Frolov M.P. Efficient IR-laser based on ZnSe:Fe crystal with continuous tuning in 3.77-4.40 μm spectral region. Quantum Electronics, V. 34, № 10, P. 879 – 988, 2004 (in Russian)
3. Zagoruiko Yu.A., Fedorenko O.A., Kovalenko N.O., Chugai O.N., Rom M.A., Mateichenko P.V. Structure and physical properties of ZnMgSe single crystals// Proceedings of SPIE, Materials and Electronics for High-Speed and Infrared Detectors.-1999.-3794.-P.96-104.
4. Yu.A. Zagoruiko, V.M.Puzikov, O.A., Fedorenko, and N.O.Kovalenko. Modification of Physical Properties of II-VI Semiconductors.- Khar'kov,: Inst. Monokristallov, 2005.- 350 p.
5. Yu.A. Zagoruiko, N.O.Kovalenko, O.A., Fedorenko. $Zn_{1-x}Mg_xSe$ single crystals: a functional material for optoelectronics. Functional Materials, V. 12, №4, P.731-734, 2005.
6. Zagoruiko Yu.A., Kovalenko N.O., Fedorenko O.O. Optical material based on single crystals of $Zn_{1-x}Mg_xSe$ solid solution / Patent of Ukraine 46429 A, 15.05.2002.